%% file: main.tex
\documentclass[10pt,conference]{IEEEtran}
\usepackage{cite}
\usepackage{amsmath,amssymb,amsfonts}
\usepackage{graphicx}
\usepackage{textcomp}
\usepackage{xcolor}
\usepackage{booktabs}
\usepackage{url}
\usepackage{pgfplots}
\usepackage{svg}
\usepackage[most]{tcolorbox}
\usepackage{hyperref}

\usepackage{microtype}
\AtBeginDocument{%
  \setlength{\abovedisplayskip}{5pt plus 2pt minus 2pt}%
  \setlength{\belowdisplayskip}{5pt plus 2pt minus 2pt}%
}

\newtcolorbox{rqbox}{
  colback=black!4,
  colframe=black!70,
  boxrule=0.4pt,
  leftrule=2.5pt,
  arc=1pt,
  boxsep=2pt,
  left=4pt,
  right=4pt,
  top=2.5pt,
  bottom=2.5pt,
  before skip=5pt,
  after skip=3pt
}
\newcommand{\rqanswer}[2]{%
  \begin{rqbox}\noindent\textbf{Answer to #1:} #2\end{rqbox}%
}

\pgfplotsset{compat=1.18}
\begin{document}
\raggedbottom
\title{Continue or Restart? The Limits of Early Failure Prediction in Coding Agents}

\author{
    \IEEEauthorblockN{Srinath Srinivasan}
    \IEEEauthorblockA{NC State University\\
    Raleigh, NC, USA\\
    ssrini27@ncsu.edu}
    \and
    \IEEEauthorblockN{Lunxiao Li}
    \IEEEauthorblockA{NC State University\\
    Raleigh, NC, USA\\
    lli66@ncsu.edu}
    \and
    \IEEEauthorblockN{Tim Menzies}
    \IEEEauthorblockA{NC State University\\
    Raleigh, NC, USA\\
    timm@ieee.org}
    \and
    \IEEEauthorblockN{Wesley Assunção}
    \IEEEauthorblockA{NC State University\\
    Raleigh, NC, USA\\
    wguezas@ncsu.edu}
}
\maketitle
\begin{abstract}
Early stopping promises to reduce the computational cost of autonomous coding agents by terminating unpromising runs before completion. Evaluating when to abandon an active run, however, requires separating the baseline difficulty of a software issue from the trajectory of an individual attempt. We study this distinction across 147,002 execution traces from two public corpora spanning four model-framework settings. Comparing pre-execution issue context against early execution prefixes, we evaluate linear count models, initial-turn text embeddings, task-conditional rankers, and a fine-tuned Qwen3-0.6B monitor. Although early prefixes modestly improve cross-task outcome ranking, within-task discrimination remains near chance across early landmarks (5k-20k tokens): initial-turn embeddings and 20k-token task-conditional models yield AUCs around 0.50, and three-seed neural evaluations across 32,406 attempts at 20k tokens achieve a mean within-task AUC of 0.50 [0.48, 0.51] across tasks of varying solve rates. To evaluate how these predictions translate into stopping decisions, we replay score-guided and matched random restarts under shared token budgets alongside uninterrupted continuation. Under a 75k-token cap, learned restarting improves resolution over matched random restarting by up to 1.98 percentage points. An exact algebraic decomposition resolves this paradox: the gain over random restarting stems almost entirely from between-task stopping allocation (avoiding stops on easy issues), whereas attempt-level selection within a task contributes near zero (-0.11 to +0.24 percentage points). Across all four settings at the 75k cap (and 216 of 224 tested budget-threshold configurations overall), uninterrupted continuation achieves higher solve rates at lower mean token cost than both restart policies. Practical evaluations of early stopping must therefore establish attempt-level discrimination and demonstrate net compute savings relative to continuation.
\end{abstract}

\begin{IEEEkeywords}
coding agents, early failure prediction, early stopping, restart policies, mining execution traces, empirical study
\end{IEEEkeywords}

\section{Introduction}
Consider a coding agent that has consumed 20,000 (20k) transcript tokens investigating a repository issue. By this point, the agent has inspected repository files and invoked initial test or edit commands, yet whether its trajectory will ultimately resolve the issue remains unknown. Suppose a failure monitor assigns the active attempt a low probability of success. Under a 75,000 (75k) token budget, aborting the run to start a fresh attempt leaves only 55k tokens for the replacement and discards all prior repository exploration. Whenever an aborted run would have succeeded within the full budget, or whenever its replacement exhausts the residual 55k budget before finishing, restarting degrades overall resolution. Evaluating early stopping therefore requires determining whether early trajectory signals justify abandoning an active attempt.

Autonomous coding agents pair language models with execution scaffolds to navigate codebases, modify source files, and run verification tests~\cite{swebench,sweagent,openhands}. Throughout this study, a \emph{task} denotes a repository issue, and a \emph{run} (or \emph{attempt}) denotes a single execution trajectory on that task. At any decision landmark, the \emph{prefix} comprises the interaction sequence of assistant turns and tool observations recorded up to that transcript-token count, prior to final patch evaluation.

Evaluating early failure predictors presents two methodological challenges. First, baseline task solvability confounds attempt-level progress. If an agent consistently resolves one issue and fails another, a monitor that estimates static task difficulty will achieve strong corpus-wide ranking even if it assigns identical scores to every attempt on the same issue. We therefore separate \emph{pooled prediction}, which ranks outcomes across all eligible runs, from \emph{within-task discrimination}, which compares successful and failed attempts on the exact same task using area under the ROC curve ($\mathrm{AUC}_{\mathrm{within}}$, where $0.50$ indicates random ranking).

Second, predictive ranking does not directly establish the resource efficiency of a stopping policy. Mid-execution interventions can follow two distinct regimes: \emph{abort-only termination}, which abandons low-scoring runs without replacement to save tokens at the expense of forfeiting any late resolutions, and \emph{restart-based stopping}, which terminates a stalled attempt to reallocate the remaining budget to a fresh attempt on the same issue. If an early monitor captures only static issue difficulty ($\mathrm{AUC}_{\mathrm{within}} \approx 0.50$), abort-only termination at $20\text{k}$ tokens merely executes a delayed task filter after burning $20\text{k}$ tokens (which is strictly dominated by pre-execution filtering at zero token cost), so we focus our policy evaluation on whether trajectory-based stopping improves resolution via cold restarts. Aborting an active run to restart improves outcomes only when a replacement attempt is more likely to solve the task within the residual budget than the original run is via continuation. Stopping rules must therefore be evaluated against both \emph{matched random restarting} (aborting the same fraction of active runs without using predictions) and uninterrupted \emph{continuation} under a shared token cap.

While prior work proposes monitors based on token probabilities, behavioral states, internal activations, and state-preserving recovery~\cite{agentstop,automata,horizons,failfast}, recent studies show that task identity confounds pooled trajectory metrics~\cite{beyond,disentangling} and that continuation often outperforms cold restarts in mathematical reasoning~\cite{problempath}. However, prior work has not tested whether task-conditional training or end-to-end neural monitors recover attempt-level signal, nor explained why score-guided policies still beat random restarts in replay. We mine $147,002$ released execution traces from Nebius SWE-rebench and NVIDIA Open-SWE-Traces~\cite{nebius,nvidia} across four model-framework combinations (\emph{settings}). Across landmarks from $5\text{k}$ to $60\text{k}$ tokens (focusing on $20\text{k}$, where nearly all runs remain active), we compare pre-execution issue features against early prefixes using action-count regressions, frozen text embeddings, task-conditional rankers, and a fine-tuned Qwen3-0.6B monitor adapted from prior work~\cite{failfast}. Using recorded-attempt replay under shared token caps, we answer three research questions:
\begin{enumerate}
\item[\textbf{RQ1.}] How much does observing an early execution prefix improve outcome prediction beyond pre-execution task context?
\item[\textbf{RQ2.}] Do early execution prefixes distinguish successful from failed attempts on the same task?
\item[\textbf{RQ3.}] What drives the performance difference between learned and random restarting, and how do both policies compare with uninterrupted continuation under a shared token budget?
\end{enumerate}

Across all four settings, we make three main contributions. First, we show via unseen-repository cross-validation, permutation controls, and equivalence testing (TOST) that while static context supports cross-task ranking, early prefixes yield near-chance within-task discrimination ($\mathrm{AUC}_{\mathrm{within}} \approx 0.50$) across $5\text{k}$ embeddings as well as $20\text{k}$ count models, task-conditional rankers, and the three-seed neural monitor, even though $87.51\%$ of primary-corpus runs have already invoked tests and $32.80\%$ have edited code by $20\text{k}$ tokens. Second, in budget-constrained replay under a $75\text{k}$ cap, we derive an exact decomposition proving that learned stopping's gain over matched random restarting (up to $1.98$ percentage points) stems from between-task stopping allocation rather than within-task attempt selection. Third, we show across $224$ budget-threshold configurations that uninterrupted continuation achieves higher solve rates at lower mean token cost than both restart policies across the $75\text{k}$ and $100\text{k}$ frontiers ($111$ of $112$ configurations, and $216$ of $224$ overall).

\section{Related Work}

\paragraph{Mining coding-agent execution traces}
Benchmarks such as SWE-bench evaluate autonomous agents on real-world repository issues~\cite{swebench}, using execution frameworks like SWE-agent~\cite{sweagent} and OpenHands~\cite{openhands} to orchestrate repository exploration and patch verification. Large-scale public corpora now capture complete execution traces from these workflows~\cite{nebius,nvidia}. These datasets extend classical mining of developer interaction histories~\cite{robbes2026} to autonomous software engineering, recording fine-grained sequences of tool invocations and environment observations. Recent empirical studies mining these traces show that issue-level properties strongly shape trajectory metrics. Mehtiyev and Assun\c{c}\~ao demonstrate that the apparent relationship between trajectory length and failure shifts substantially once task identity is controlled~\cite{beyond}. Complementary analyses examine how failure cascades develop during execution~\cite{process} and whether behavioral features carry consistent predictive weights across issues and repeated trials~\cite{consistency}. These findings motivate mining multi-attempt corpora to test whether early trajectory observations capture attempt-specific progress beyond static issue difficulty.

\paragraph{Early failure prediction in agent runs}
Prior work proposes diverse signals to anticipate unsuccessful agent runs before completion. Trajectory-based approaches mine historical execution experience to terminate unpromising patch generation~\cite{eet} or model agent behavior through state transitions and late-emerging failure indicators~\cite{automata,sparse}. A second line of work monitors execution sequentially by estimating residual resource requirements~\cite{bagen} or applying sequential statistical testing to intermediate verifier scores~\cite{evaluator}. A third family extracts signals directly from the underlying language model, using generated token probabilities~\cite{agentstop} or probing internal representations to predict edit outcomes and likely failures~\cite{doomed,horizons}. Whereas existing monitors are primarily evaluated on pooled benchmark accuracy or total token reduction, we evaluate whether early prefixes rank competing attempts on the exact same software issue.

\paragraph{Restart strategies and state recovery}
Restart policies originate in randomized algorithms, where Luby et al.\ established optimal and universal restart schedules for heavy-tailed runtime distributions based on elapsed execution time~\cite{luby1993}. In autonomous coding workflows, mid-episode interventions can either start fresh or transfer accumulated context. Scrouting uses an initial repository exploration phase to route the issue and its gathered context to a downstream solver~\cite{scrouting}. \textit{FailFast--RestartSmart} pairs a 0.6B failure monitor, trained on terminal outcomes and intermediate test supervision, with a recovery mechanism that retains code edits from aborted runs~\cite{failfast}. Across four coding-agent policies, their system reports 14.6--20.4\% execution-token savings at a 5\% false-positive target. Notably, at a 25\% false-positive target with Qwen3.6-27B, transferring saved edits raises resolution from 66.6\% to 71.8\%, whereas restarting cold reaches only 66.8\%. That sharp gap between state-preserving recovery and cold restarting highlights the cost of discarding prior exploration. Building on this insight, we adapt their 0.6B terminal-outcome monitor to evaluate cold replacement attempts under shared token budgets and isolate how sunk exploration tokens affect the restart tradeoff.

\paragraph{Disentangling task difficulty from stopping policies}
Two concurrent studies examine the gap between problem difficulty and attempt-level intervention. Investigating the Nebius OpenHands corpus~\cite{nebius} alongside additional benchmarks, EsfandyariDoulabi et al.\ show that high pooled AUROC largely reflects cross-task difficulty even when within-task discrimination remains near chance, concluding that cross-task budget allocation is more effective than abort-only stopping in this regime~\cite{disentangling}. In mathematical reasoning, Bulut similarly finds that early prefix predictions offer minimal within-problem discrimination and that continuing an active trace outperforms cold restarts under matched token budgets~\cite{problempath}.

Our work extends this emerging perspective across four coding-agent settings by connecting within-task evaluation directly to the mechanics of restart policies. First, we evaluate whether task-conditional training or a fine-tuned neural monitor recovers attempt-level signal that count features and frozen embeddings miss, including across tasks stratified by empirical solvability. Second, we explain the apparent paradox in which a learned stopping rule beats a matched random restart baseline despite near-chance within-task ranking. By decomposing replay outcomes into between-task and within-task selection and accounting for both replacement credit and forfeited original successes, we show how task selection drives the gain over random restarting while uninterrupted continuation dominates both policies across the shared-budget frontier.

\section{Data and Method}
\label{sec:method}
Our methodology comprises four stages: curating multi-attempt execution traces across four model-framework settings, extracting prefix representations at fixed token landmarks, evaluating cross-task and within-task predictive discrimination, and replaying stopping policies under shared token budgets.

\subsection{Released runs and labels}
\label{sec:data}
We analyze 147,002 execution traces drawn from two public corpora of autonomous coding agents solving benchmark repository issues. Each trace represents a single \emph{run} (or \emph{attempt}) $i$ by an agent assigned to resolve a specific repository issue, which we term a \emph{task} $t$, belonging to a source code \emph{repository}. We define a \emph{setting} as a unique combination of an underlying large language model and an agent execution framework.

Our primary corpus, released by Nebius~\cite{nebius}, executes OpenHands with Qwen3-Coder-480B-A35B-Instruct~\cite{qwen3coder} on SWE-rebench tasks~\cite{swerebench}. To evaluate generalization across architectures and scaffolding, we incorporate three slices from NVIDIA Open-SWE-Traces~\cite{nvidia}: SWE-agent with Qwen3.5-122B-A10B~\cite{qwen35}, OpenHands with Qwen3.5-122B-A10B, and OpenHands with MiniMax-M2.5~\cite{minimaxm25}. Table~\ref{tab:data} summarizes the corpus statistics across all four settings. Because individual benchmark issues can appear in multiple corpora, task counts overlap across settings.

Each completed run terminates with a binary \emph{outcome label} $Y_i \in \{0, 1\}$ produced by the benchmark evaluation harness, which applies the agent's final code patch to the repository and executes the verification test suite. We set $Y_i = 1$ if the patch resolves the issue and $Y_i = 0$ if the issue remains unresolved, excluding unlabelled records from our analysis.
\input{tables/data.tex}

\subsection{Trace representation and decision landmarks}
\label{sec:landmarks}
During execution, an agent interacts with its environment through an alternating sequence of assistant turns and tool observations. To establish a uniform measure of computational progress within each setting, we measure trajectory length in \emph{transcript tokens}, counted across all recorded messages using that setting's native model tokenizer, and denote the total transcript length of completed run $i$ by $L_i$. We also normalize raw tool invocations into standard behavioral categories covering file inspection, codebase search, code editing, and test execution.

A \emph{decision landmark} $b$ is a fixed transcript-token threshold at which a monitor evaluates an ongoing run. At landmark $b$, the cohort of \emph{eligible} (or \emph{active}) runs consists exclusively of attempts whose total token length exceeds the threshold ($L_i > b$), ensuring that a stopping intervention remains actionable. For each eligible run, the predictor observes only the execution \emph{prefix} completed up to token $b$. We evaluate predictions across seven landmarks ($b \in \{5\text{k}, 10\text{k}, 15\text{k}, 20\text{k}, 30\text{k}, 40\text{k}, 60\text{k}\}$ tokens). Because shorter runs terminate as $b$ increases, we report the \emph{surviving fraction} (the proportion of initial runs satisfying $L_i > b$) at each landmark to characterize the active population. We designate $b = 20\text{k}$ tokens as our primary decision point because agents have already engaged in substantial environment interaction by this stage while nearly all runs remain active (Fig.~\ref{fig:landmarks}, right). On the primary corpus at $20\text{k}$ tokens, $87.51\%$ of active runs have already executed a detected test command and $32.80\%$ have a detected edit action.

\subsection{Evaluation metrics and task clustering}
\label{sec:eval}
Before assessing individual predictors, we quantify the degree to which trial outcomes cluster by task using the \emph{intraclass correlation} (ICC), computed via an analysis-of-variance estimator for binary outcomes~\cite{ridout1999} across tasks with repeated attempts. Values of ICC approaching $1.0$ indicate that attempts on the same issue predominantly share the same terminal outcome.

At a given landmark $b$, a failure monitor assigns each eligible run $i$ a scalar \emph{success score} $s_i \in [0, 1]$, where higher values indicate a higher predicted probability of resolving the task. We evaluate the calibration and sharpness of probability estimates using \emph{log loss}~\cite{gneiting2007strictly} (the mean negative log-likelihood of the true outcome $Y_i$, where lower values reflect superior probabilistic accuracy) and evaluate ranking discrimination using two formulations of the area under the ROC curve (AUC)~\cite{hanley1982}:

\noindent\textbf{1) Pooled AUC} pairs every successful eligible run with every failed eligible run across the entire setting, computing the fraction of pairs in which the successful run receives a higher score (with half credit for ties, where $0.50$ represents random ranking). Because pooled pairs span different issues, pooled AUC captures both cross-task difficulty and attempt-level differences.

\noindent\textbf{2) Within-task AUC} ($\mathrm{AUC}_{\mathrm{within}}$) isolates attempt-level discrimination by restricting comparisons to pairs of runs executed on the exact same task. We define a \emph{mixed-outcome task} at landmark $b$ as a task whose active runs include at least one success and at least one failure. Let $P_t$ and $F_t$ denote the sets of successful and failed active runs on task $t$, respectively. Aggregating across all mixed-outcome tasks, within-task AUC is defined as:
\begin{equation}
\mathrm{AUC}_{\mathrm{within}} =
\frac{\sum_t \sum_{i\in P_t,\, j\in F_t}
\!\left[\mathbf{1}(s_i>s_j)+\tfrac{1}{2}\mathbf{1}(s_i=s_j)\right]}
{\sum_t |P_t|\,|F_t|},
\label{eq:auc-within}
\end{equation}
where $\mathbf{1}(\cdot)$ is the indicator function. Each valid within-task pair carries equal weight in Eq.~\eqref{eq:auc-within}; for instance, a mixed-outcome task with two active successes and three active failures contributes $|P_t|\,|F_t| = 6$ pairs.

To account for hierarchical dependence in our data during uncertainty estimation, we compute 95\% cluster-bootstrap confidence intervals that resample whole clustering units with replacement~\cite{field2007} from saved out-of-fold predictions. Specifically, pooled and replay metrics resample repositories (and training seeds for the neural monitor), whereas within-task AUC resamples tasks.

\subsection{Prediction models for RQ1 and RQ2}
\label{sec:models}
To measure how much early execution trajectories improve outcome prediction over static issue descriptions (RQ1) and whether they distinguish competing attempts on the same task (RQ2), we train linear and conditional models using five-fold \emph{unseen-repository cross-validation}. In this splitting protocol, all tasks belonging to a given repository are assigned to the same fold, ensuring that models are evaluated exclusively on repositories held out during training.

Our \emph{issue baseline} extracts eight pre-execution count features from the problem statement covering text length, code blocks, error and reproduction indicators, URLs, and referenced file paths. Our \emph{issue-plus-prefix model} augments these eight issue features with 18 behavioral count features extracted from the observed prefix up to landmark $b$, capturing distinct files inspected, repeated shell commands, error outputs, cumulative text volume, code edits, and test invocations (two fine-grained test-output features are specific to the primary corpus trace schema, yielding 16 prefix features for the Open-SWE settings). Both models use L2-regularized logistic regression fit on log-transformed, standardized feature counts. Comparing the issue-plus-prefix model against the issue baseline isolates the predictive gain contributed by observing execution.

Because pooled logistic regression fits weights to variation both across tasks and within tasks, we also train a \emph{task-conditional logistic model}~\cite{breslow1980statistical} restricted to mixed-outcome tasks. By conditioning on task identity, this formulation eliminates the task-level intercept and estimates feature weights solely from differences between successful and failed attempts on the same issue using prefix features. We also evaluate cross-setting transfer at $20\text{k}$ tokens by applying the $16$-feature conditional model trained on one setting directly to another after holding out overlapping repositories and tasks. To verify that our conditional estimation pipeline remains unbiased under the null, we validate the estimator against a permutation control that shuffles outcome labels within each task. To quantify how tightly within-task AUC concentrates around chance ($0.50$), we evaluate equivalence bounds of $[0.45, 0.55]$ and $[0.47, 0.53]$ using two one-sided tests (TOST) at the $\alpha = 0.05$ level via 90\% task-bootstrap confidence intervals~\cite{schuirmann1987,lakens2017}.

Beyond standard count regressions, we evaluate three supplementary designs to test alternative sources of predictive signal:

\paragraph{Task metadata and solvability stratification}
We test whether augmenting count models with Open-SWE-Traces task categories (\texttt{bug fix}, \texttt{enhancement}, \texttt{feature request}, \texttt{other})~\cite{nvidia} and programming language under unseen-repository cross-validation improves cross-task and within-task discrimination. To test whether attempt-level discrimination is stronger on easier issues, we also stratify mixed-outcome tasks at the 20k landmark into low ($<25\%$), medium ($25\text{--}75\%$), and high ($>75\%$) empirical solve-rate tiers (Table~\ref{tab:pass-strata}).

\paragraph{Frozen text embeddings}
To test whether dense lexical representations capture early trajectory signals missed by summary counts, we encode the truncated issue description, the initial assistant turn, and the first tool observation on the primary corpus using the frozen \texttt{all-MiniLM-L6-v2} sentence encoder~\cite{sbert,minilm}, concatenating these embeddings with initial prompt-context metadata as inputs to a regularized linear classifier.

\paragraph{Task-rate oracle}
To establish an upper bound on task-level predictability and test whether prefix features add value when baseline task solvability is already known, we construct a diagnostic \emph{task-rate oracle} on tasks with at least four active attempts. Across repeated random splits, we hold out half of a task's attempts to compute its empirical solve fraction and supply that fraction as a feature when scoring the remaining half, both standalone and combined with prefix counts.

\subsection{Adapted neural monitor}
\label{sec:neural}
To test whether end-to-end representation learning extracts fine-grained within-task trajectory cues that count models and frozen embeddings miss (RQ2), we adapt the neural failure monitor architecture from FailFast--RestartSmart~\cite{failfast}. Specifically, we fine-tune Qwen3-0.6B~\cite{qwen3} using low-rank adaptation (LoRA)~\cite{lora} and a linear classification head to predict terminal task resolution at the 20k-token landmark under a $4,096$-token input limit. We compare two input configurations trained and evaluated on identical data splits: an \emph{issue-only arm} that receives the problem description, and an \emph{issue-plus-prefix arm} that receives the problem description alongside the most recent transcript messages completed by $20\text{k}$ tokens. When present in the preceding prefix, the issue-plus-prefix input also includes the latest earlier message matching an editor-tool heuristic (which captures editor-tool invocations including file-viewing and modification operations, and is simply omitted when no prior message matches).

To prevent static task difficulty from dominating gradient updates and isolate attempt-level discrimination for RQ2, each optimization step samples a matched pair consisting of one successful and one failed attempt from the same mixed-outcome task, minimizing a composite objective that combines binary cross-entropy (BCE) loss with a pairwise ranking loss~\cite{sculley2010combined} that rewards assigning a higher score to the successful attempt. By construction, balanced within-task pair sampling fixes the training prevalence of every task at $0.50$. For the issue-only arm, where both attempts in a pair share identical input text, the BCE objective has an analytical minimum at a constant prediction of $0.50$ (yielding a baseline log loss of $\ln 2 \approx 0.693$ and $\mathrm{AUC}_{\mathrm{within}} = 0.500$ because all attempts on a task receive identical scores), providing a null reference for measuring attempt-level signal in the issue-plus-prefix arm.

We conduct our neural evaluation across all primary-corpus tasks with at least 11 active attempts at the 20k landmark, sampling 11 attempts per task uniformly without outcome stratification to form a cohort of 32,406 attempts across 2,946 tasks. We partition this cohort using three-fold unseen-repository cross-validation so that every task is scored out-of-fold, and we train three random seeds per fold with checkpoint selection governed by validation within-task AUC. Because the three seeds evaluate on identical held-out repository folds, we report mean metrics aggregated across seeds and construct joint repository-and-seed bootstrap intervals.

\subsection{Budget-constrained restart replay for RQ3}
\label{sec:replay}
To determine how early failure scores translate into operational stopping decisions and resource consumption (RQ3), we perform \emph{recorded-attempt replay} across all tasks with at least two recorded attempts. In each replay trial, an initial attempt $i$ executes until landmark $b = 20\text{k}$. If a \emph{stopping policy} $\pi$ terminates run $i$ at landmark $b$, the system initiates a cold replacement attempt $j$ selected uniformly at random from the remaining recorded attempts on the same task, and we compute expected outcomes over all valid replacement choices while giving each starting attempt equal weight. Using out-of-fold success scores $s_i$ from the pooled issue-plus-prefix count logistic regression model (which spans all four settings and reflects the combined task and prefix signal deployed in standard monitors), we compare three primary policies: (1)~\emph{Learned restart ($L$)}, which terminates an active run at landmark $b$ if its out-of-fold success score $s_i$ falls below a fixed percentile threshold (evaluated across seven thresholds spanning the 2nd to 98th score percentiles); (2)~\emph{Matched random restart ($R$)}, which terminates the exact same overall fraction of active runs at landmark $b$ as the corresponding learned threshold, selecting runs uniformly at random without consulting $s_i$; and (3)~\emph{Continuation}, which permits the initial attempt $i$ to proceed to completion without intervention. (Terminating active runs at landmark $b$ without replacement, fixed truncation, corresponds to setting replacement credit $R_j \equiv 0$, which upper-bounds solve rate by the pre-landmark completion rate and cannot improve resolution over continuation.)

We first evaluate policies in an \emph{uncapped replay} regime where initial and replacement runs execute until their recorded completion lengths. To model realistic compute constraints, we then enforce a shared hard \emph{token cap} $B \in \{30\text{k}, 50\text{k}, 75\text{k}, 100\text{k}\}$ governing the combined token expenditure of the initial run and any subsequent replacement. When a policy aborts run $i$ at landmark $b$, the replacement attempt $j$ (with recorded length $L_j$ and outcome $Y_j \in \{0, 1\}$) receives a \emph{residual token budget} of $B - b$. Its resulting \emph{replacement success credit} $R_j$ and combined \emph{transcript-token cost} $K_j$ are given by:
\begin{align}
R_j &= Y_j\,\mathbf{1}(L_j\leq B-b), \label{eq:credit}\\
K_j &= b+\min(L_j,B-b). \label{eq:cost}
\end{align}
Under Eqs.~\eqref{eq:credit}--\eqref{eq:cost}, a replacement attempt earns resolution credit only if it solves the issue and completes within the remaining $B - b$ tokens. For example, stopping at $b = 20\text{k}$ under a $B = 75\text{k}$ cap leaves $55\text{k}$ tokens: a successful replacement requiring $50\text{k}$ tokens earns full credit ($R_j = 1$) at a total cost of $K_j = 70\text{k}$ tokens, whereas a replacement requiring $60\text{k}$ tokens exhausts the cap at $K_j = 75\text{k}$ and earns $R_j = 0$. A policy's overall \emph{solve rate} $V(\pi)$ is the mean resolution credit achieved across all starting attempts and replacement draws.

\paragraph{Decomposing learned versus random stopping}
When a learned restart policy achieves a higher solve rate than matched random restart ($V(L) > V(R)$), that advantage can arise from two distinct mechanisms: \emph{between-task selection} (concentrating stops on tasks where restarting yields a more favorable expected return) and \emph{within-task selection} (preferentially aborting the weaker attempts on a given task). To isolate how much of the policy difference stems from attempt-level failure detection~\cite{kitagawa1955components}, we derive an exact algebraic decomposition of $V(L) - V(R)$.

Under a fixed cap $B$, let $C_i = Y_i\mathbf{1}(L_i \leq B)$ indicate whether continuing starting attempt $i$ resolves the task within budget, and let $G_i$ denote the expected replacement credit obtained by averaging $R_j$ in Eq.~\eqref{eq:credit} over all other recorded attempts $j$ on the same task. Defining $v_i = G_i - C_i$ as the net expected solve-rate gain of restarting attempt $i$, and letting $S_i^{\pi} \in [0, 1]$ denote policy $\pi$'s stopping indicator (or stopping probability under random restart, with $S_i^{\pi} = 0$ for runs that finished prior to $b$), the replay solve rate across all $N$ starting attempts is:
\begin{equation}
 V(\pi)=\frac{1}{N}\sum_{i=1}^{N}\left[C_i+S_i^{\pi}(G_i-C_i)\right].
 \label{eq:policy-value}
\end{equation}
Equation~\eqref{eq:policy-value} shows that the net effect of stopping attempt $i$ equals the replacement success credited ($S_i^{\pi} G_i$) minus the original continuation success forfeited ($S_i^{\pi} C_i$). Let $D_i = S_i^{L} - S_i^{R}$ denote the stopping differential between the learned and matched random policies on attempt $i$, and let $A_t = \{i \in t : L_i > b\}$ denote the set of $n_t = |A_t|$ active starting attempts on task $t$ at landmark $b$, with $\bar{D}_t$ and $\bar{v}_t$ denoting the task-level means of $D_i$ and $v_i$ over $A_t$. The solve-rate differential $V(L) - V(R)$ decomposes exactly into:
\begin{align}
\Delta_{\mathrm{between}} &=\sum_{t:\,n_t>0}\frac{n_t}{N}\bar D_t\bar v_t, \label{eq:decomp-between}\\
\Delta_{\mathrm{within}} &=\frac{1}{N}\sum_{t:\,n_t>0}\sum_{i\in A_t}(D_i-\bar D_t)(v_i-\bar v_t). \label{eq:decomp-within}
\end{align}
Here, $\Delta_{\mathrm{between}}$ quantifies the solve-rate difference attributable to task-level stopping rates ($\bar{D}_t$), whereas $\Delta_{\mathrm{within}}$ captures the covariance between policy stopping decisions and attempt-level restart value within the same task.

\paragraph{Cost-solve efficiency frontier}
To evaluate practical resource efficiency at the 20k landmark, we compare the mean solve rate and mean transcript-token cost of every policy across all four settings, four token caps, and seven stopping thresholds. Within a given setting and token cap $B$, a policy configuration is \emph{Pareto-dominated} if an alternative policy achieves at least as high a solve rate at less than or equal mean token cost, with strict improvement on at least one dimension. Mapping dominance across the complete 224-configuration grid characterizes the empirical efficiency frontier of cold restarts relative to uninterrupted continuation.

\section{Results}

\subsection{Baseline task clustering across settings}
Across all four settings, terminal outcomes cluster strongly by task, with observed-scale intraclass correlation (ICC) reaching $0.71$ or higher in every corpus (Table~\ref{tab:data}). Consequently, repeated attempts on the same repository issue predominantly share the same outcome.

Comparing the two Qwen3.5 settings across their $7,102$ shared tasks further illustrates how intrinsic issue difficulty persists across agent scaffolding. The empirical task solve rates of SWE-agent and OpenHands exhibit a Spearman rank correlation of $0.67$, confirming that issues that are more tractable under one framework remain more tractable under the other. Weighting each shared task equally, SWE-agent achieves a mean solve rate of $47.30\%$ compared to $33.24\%$ for OpenHands, yielding a framework difference of $14.06$ percentage points ($95\%$ repository-bootstrap CI: $[12.26, 15.95]$). Together, high within-task outcome clustering and cross-framework rank agreement establish that static task properties account for a substantial share of outcome variation.

\subsection{RQ1: Value of observing an early prefix}
To evaluate whether early execution trajectories improve outcome prediction beyond pre-execution context (RQ1), we compare pooled performance at early landmarks across models trained on the empirical task distribution. At $5\text{k}$ tokens on the primary corpus, augmenting the issue baseline with prefix action counts increases pooled AUC from $0.56$ to $0.58$, while log loss remains nearly unchanged ($0.69$ in both arms). By $20\text{k}$ tokens, prefix action counts increase pooled AUC from $0.57$ to $0.61$, reflecting that cumulative early interaction rates correlate with underlying issue difficulty across tasks. For frozen text embeddings of the initial turn at $5\text{k}$ tokens, pooled AUC remains at $0.62$ in both the issue-only and issue-plus-prefix arms (with the paired gain interval spanning zero and log loss remaining at $0.67$).

Supplementary checks confirm that static task context drives this cross-task ranking while leaving attempt-level discrimination unchanged. Even when restricting comparisons to pairs of runs from different tasks within the same repository on the primary corpus, eliminating repository-level confounding, issue-plus-prefix scores rank cross-task pairs with an AUC of $0.56$ $[0.55, 0.59]$ at $20\text{k}$ tokens ($0.62$ $[0.59, 0.66]$ at $40\text{k}$), compared to $0.49$ ($0.54$ at $40\text{k}$) for attempts on the exact same task. Similarly, adding Open-SWE task categories under unseen-repository cross-validation increases issue-baseline pooled AUC by approximately $0.02$ across landmarks while leaving within-task AUC unchanged ($\Delta\mathrm{AUC}_{\mathrm{within}} < 0.01$, with no statistically significant gains across task-category or language subgroups after Holm--Bonferroni multiplicity correction~\cite{holm1979}).

\rqanswer{RQ1}{Early action counts and static metadata yield modest gains in cross-task pooled ranking ($0.02\text{--}0.04$ AUC), whereas initial-turn text embeddings provide no consistent pooled improvement over pre-execution issue text alone.}

\subsection{RQ2: Ranking attempts at the same task}

\subsubsection{Early action counts and frozen embeddings}
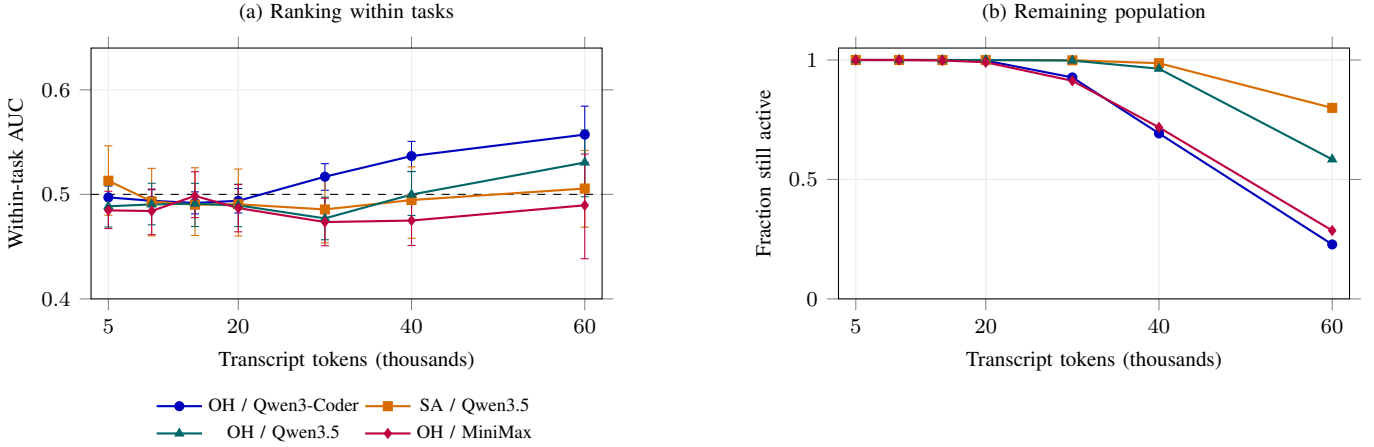
\begin{figure*}[t]
\centering
\input{figures/landmarks.tex}
\caption{Prediction across fixed token landmarks. Left: within-task AUC for issue-plus-prefix count models, with 95\% task-bootstrap confidence intervals at each landmark. The dashed line indicates random ranking (0.50). Right: surviving fraction of initial runs still active at each landmark. OH: OpenHands; SA: SWE-agent.}
\label{fig:landmarks}
\end{figure*}

We next evaluate whether prefix representations separate successful from failed attempts on the same issue. At $5\text{k}$ tokens on the primary corpus, the issue-plus-prefix count model achieves a within-task AUC of $0.50$ $[0.49, 0.51]$, indicating chance-level attempt discrimination despite the pooled gain observed in RQ1. At the primary $20\text{k}$ decision point, within-task AUC remains near $0.49\text{--}0.50$ across all four settings (Fig.~\ref{fig:landmarks}, left). At intermediate landmarks of $30\text{k}$ and $40\text{k}$ tokens, within-task AUC remains indistinguishable from chance across all three Open-SWE settings and rises only modestly on the primary corpus to $0.52$ $[0.51, 0.53]$ at $30\text{k}$ ($93\%$ of runs active) and $0.54$ $[0.53, 0.55]$ at $40\text{k}$ ($69\%$ active), a stage at which aborting a run forfeits $30\text{k}\text{--}40\text{k}$ sunk tokens and compresses the residual budget available for a replacement attempt. By $60\text{k}$ tokens, within-task AUC on the primary corpus reaches $0.56$, but the surviving fraction drops to $23\%$ of initial runs (Fig.~\ref{fig:landmarks}, right), reflecting both longer observation horizons and selective attrition of shorter trajectories.

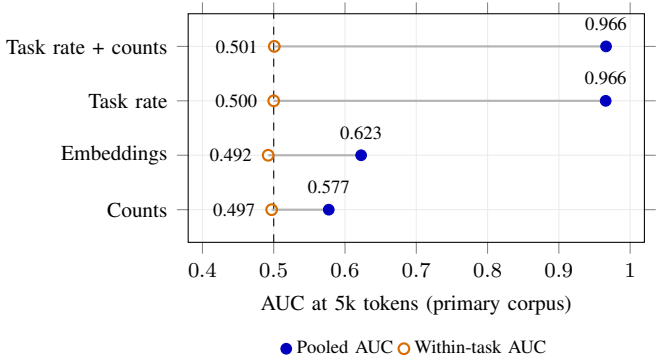
\begin{figure}[t]
\centering
\input{figures/pooled_within.tex}
\caption{Comparison of pooled AUC and within-task AUC at 5k tokens on the primary corpus. Each horizontal line connects a predictor's pooled and within-task performance. Counts: issue features plus prefix action counts. Embeddings: frozen sentence embeddings of the issue and initial step. Task rate: empirical solve fraction computed from a reserved half of each task's attempts (evaluated across tasks with at least four attempts and averaged over three random splits). The dashed line marks random ranking (0.50).}
\label{fig:pooled-within}
\end{figure}

The diagnostic task-rate oracle demonstrates how near-perfect pooled discrimination can coexist with chance-level within-task ranking. Because task solvability is heavily polarized across issues ($\mathrm{ICC} \ge 0.71$), predicting outcomes solely from the empirical solve fraction of a task's reserved attempts at $5\text{k}$ tokens achieves a mean pooled AUC of $0.97$ while yielding a within-task AUC of $0.50$ by construction. Combining prefix action counts with the task-rate oracle leaves pooled AUC at $0.97$ and yields a within-task AUC of $0.50$ at $5\text{k}$ tokens (rising to $0.53$ at $60\text{k}$ tokens), confirming that early prefix features add negligible discrimination once baseline task solvability is known. Figure~\ref{fig:pooled-within} contrasts pooled and within-task AUC across four predictors at $5\text{k}$ tokens: whereas pooled AUC spans $0.58$ to $0.97$, within-task AUC clusters tightly around $0.50$, including for frozen text embeddings ($0.49$ $[0.48, 0.50]$).

\subsubsection{Task-conditional and cross-setting training}
To test whether removing between-task variance during training uncovers attempt-level signal, we evaluate task-conditional logistic regression fit directly on mixed-outcome tasks. At $20\text{k}$ tokens on the primary corpus, the conditional model achieves a within-task AUC of $0.50$ $[0.49, 0.51]$ across $38,447$ successful-failed pairs drawn from $1,737$ mixed-outcome tasks. The conditional model's $90\%$ task-bootstrap interval $[0.49, 0.51]$ falls entirely inside both the $[0.45, 0.55]$ and $[0.47, 0.53]$ TOST equivalence bounds, while the within-task label-permutation control remains centered at chance ($0.50$ $[0.49, 0.51]$).

Across the three Open-SWE settings at $20\text{k}$ tokens, conditional within-task AUC is $0.50$ $[0.48, 0.52]$ for OpenHands/Qwen3.5, $0.51$ $[0.49, 0.53]$ for OpenHands/MiniMax-M2.5, and $0.48$ $[0.45, 0.52]$ for SWE-agent/Qwen3.5, with all three $95\%$ confidence intervals spanning $0.50$. Both OpenHands settings pass the within-task permutation control and have $90\%$ confidence intervals falling strictly within the $[0.47, 0.53]$ equivalence bound (we refrain from formal TOST equivalence claims on SWE-agent/Qwen3.5 because its permutation control deviated from $0.50$). Evaluating cross-setting transfer at $20\text{k}$ tokens, with overlapping repositories and task identifiers held out, likewise yields within-task AUCs of at most $0.50$ across all three transfer pairs (SWE-agent/Qwen3.5 to OpenHands/Qwen3.5, and the primary setting to both SWE-agent/Qwen3.5 and OpenHands/MiniMax-M2.5).

\subsubsection{Adapted neural monitor}
Table~\ref{tab:neural-large} reports the three-seed evaluation of the LoRA-adapted Qwen3-0.6B monitor across $32,406$ attempts at the $20\text{k}$ landmark. Across $16,818$ successful-failed pairs from $859$ mixed-outcome tasks, the issue-plus-prefix monitor achieves a mean within-task AUC of $0.50$ $[0.48, 0.51]$. Performance remains near $0.50$ within every individual repository fold and training seed, and weighting each mixed-outcome task equally produces identical results. As expected under balanced within-task pair sampling, the issue-only reference arm converges to the analytical baseline ($\mathrm{AUC}_{\mathrm{within}} = 0.500$, log loss $= 0.693$).

The neural monitor's $90\%$ repository-and-seed bootstrap interval $[0.48, 0.51]$ lies strictly within the $[0.45, 0.55]$ equivalence bound.

\input{tables/neural_large}

\subsubsection{Stratification by task solvability and late-trajectory signals}
To test whether early attempt-level discrimination emerges on easier issues where failures might reflect avoidable early missteps, Table~\ref{tab:pass-strata} stratifies mixed-outcome tasks at $20\text{k}$ tokens into low ($<25\%$), medium ($25\text{--}75\%$), and high ($>75\%$) empirical solve-rate tiers. Across all three tiers, within-task AUC ranges from $0.49$ to $0.51$ for the count-based model and from $0.49$ to $0.50$ for the three-seed neural monitor, with every confidence interval containing $0.50$. High-solvability tasks therefore show the same near-chance early ranking as harder tasks.

\input{tables/pass_strata.tex}

To verify that successful and failed trajectories on the same task do eventually diverge by completion, we compute an oracle end-of-run diagnostic measuring \emph{reference-hunk recall} (the fraction of ground-truth patch diff hunks overlapped by the agent's final edits, scored as $0$ when no patch is produced). At run termination, reference-hunk recall achieves a within-task AUC of $0.62$ $[0.61, 0.64]$, confirming that attempt-level divergence materializes over the full episode even though it remains undetectable from prefix representations at $20\text{k}$ tokens.

\rqanswer{RQ2}{Across initial-turn frozen embeddings (5k tokens) as well as 20k-token action-count models, task-conditional rankers, cross-setting transfers, and the three-seed Qwen3-0.6B neural evaluation ($\mathrm{AUC}_{\mathrm{within}} = 0.50$ $[0.48, 0.51]$), early prefixes provide near-chance discrimination between successful and failed attempts on the same task across all solvability tiers.}

\subsection{RQ3: Evaluating restart policies under shared token budgets}
\begin{figure}[t]
\centering
\includegraphics[width=\linewidth]{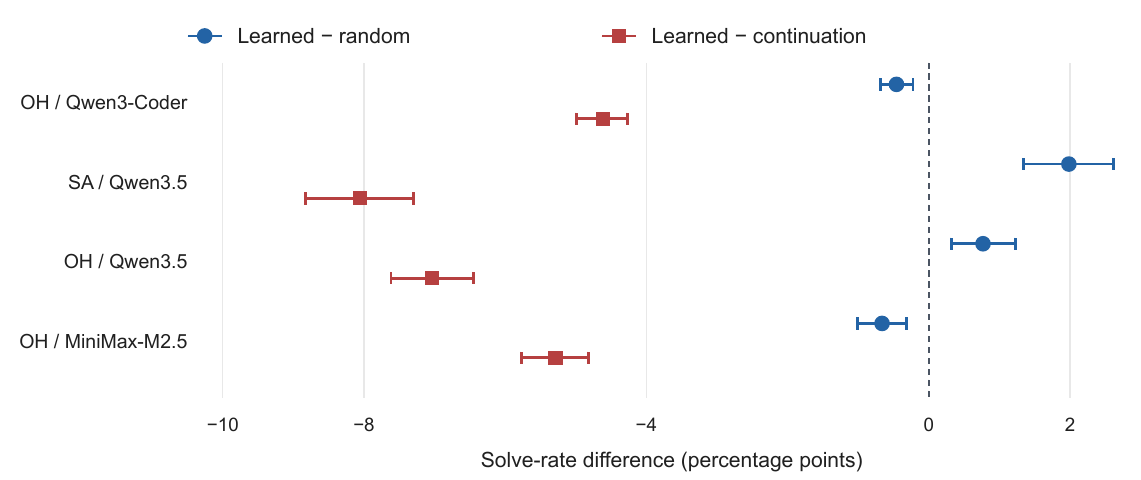}
\caption{Solve-rate difference (in percentage points) between median-threshold learned restart and matched random restart (blue), and between learned restart and uninterrupted continuation (red), evaluated at the 20k-token landmark under a shared 75k-token cap. Error bars indicate 95\% repository-bootstrap confidence intervals (5,000 draws), with positive values favoring learned restart. While learned restart outperforms random restart in the two Qwen3.5 settings, uninterrupted continuation outperforms learned restart across all four settings.}
\label{fig:budget}
\end{figure}

To understand how early failure scores behave when deployed as stopping rules (RQ3), Table~\ref{tab:restart} compares learned restart (stopping active runs scoring below the median eligible score at $20\text{k}$ tokens) against matched random restart. In the uncapped replay regime, solve-rate differences between learned and random restarting are negligible across all four settings (within $0.20$ percentage points): because unconstrained replacement attempts run to completion without budget truncation and within-task ranking is near chance, replacing an active attempt on a given task preserves the task's expected solve rate. Under a shared $75\text{k}$-token cap, however, the solve-rate differential $V(L) - V(R)$ spans $-0.66$ to $+1.98$ percentage points, favoring learned restart in both Qwen3.5 settings ($+1.98$ and $+0.77$ percentage points) while favoring random restart in the primary and MiniMax settings. Figure~\ref{fig:budget} contrasts these $75\text{k}$ learned-versus-random differentials (blue) against the corresponding comparisons with uninterrupted continuation (red).
\input{tables/restart.tex}

\subsubsection{Between-task selection drives the 75k learned-versus-random differential}
Applying the exact decomposition from Eqs.~\eqref{eq:policy-value}--\eqref{eq:decomp-within} at the median threshold explains how learned restarting can outperform matched random restarting despite near-chance within-task AUC. As shown in Table~\ref{tab:replay-decomp}, the between-task selection term $\Delta_{\mathrm{between}}$ ($-0.70$ to $+1.75$ percentage points) matches the sign of the total differential $V(L) - V(R)$ and exceeds the magnitude of the within-task term $\Delta_{\mathrm{within}}$ in every setting. By contrast, $\Delta_{\mathrm{within}}$ remains small across all four corpora ($-0.11$ to $+0.24$ percentage points) and never determines the direction of the overall policy difference.
\input{tables/replay_decomposition.tex}

The task-level net gain $\bar{v}_t = \bar{G}_t - \bar{C}_t$ in Eq.~\eqref{eq:decomp-between} clarifies why $\Delta_{\mathrm{between}}$ changes sign across settings. Because the learned policy assigns lower scores to harder tasks, it concentrates stops on tasks where both the replacement success rate within the remaining $55\text{k}$ tokens ($\bar{G}_t$) and the original continuation success rate ($\bar{C}_t$) are lower than the corpus average. Consequently, learned stopping yields less replacement credit than random stopping, but it also forfeits fewer original continuation successes. In the two Qwen3.5 settings, the reduction in forfeited continuation successes outweighs the reduction in replacement credit, producing a net positive $\Delta_{\mathrm{between}}$ purely through task-level stopping allocation.

\subsubsection{Uninterrupted continuation dominates the efficiency frontier}
Whereas comparing learned against random restart evaluates how a fixed stopping quota is distributed across runs, comparing against uninterrupted continuation tests whether cold restarting saves compute or improves resolution overall. Extending replay across all seven score-percentile thresholds, four settings, and four hard token caps yields $224$ learned and random restart configurations at the $20\text{k}$ decision point (Table~\ref{tab:frontier-summary}). Under the primary $75\text{k}$-token cap, where replacement runs receive $55\text{k}$ residual tokens, uninterrupted continuation achieves both a strictly higher solve rate and a strictly lower mean transcript-token cost than every learned and random restart configuration across all four settings ($56$ of $56$ configurations). In SWE-agent/Qwen3.5, for example, continuation resolves $28.71\%$ of starting attempts within the $75\text{k}$ cap (down from $47.30\%$ uncapped across shared tasks, as many SWE-agent runs exceed $75\text{k}$ tokens) at a mean expenditure of $68.10\text{k}$ tokens, whereas median learned restart resolves $20.66\%$ at a mean expenditure of $69.93\text{k}$ tokens. This dominance persists at the $100\text{k}$-token cap ($80\text{k}$ residual tokens), where continuation Pareto-dominates $55$ of $56$ configurations, leaving only a single boundary point for OpenHands/MiniMax ($2$nd-percentile threshold, which adds less than $0.01$ percentage points in solve rate at an added cost of $376$ mean tokens). Across all $224$ tested configurations spanning $B \in \{30\text{k}, 50\text{k}, 75\text{k}, 100\text{k}\}$, continuation Pareto-dominates $216$ configurations, with the other seven non-dominated cases being exact ties at the $30\text{k}$ cap.
\input{tables/frontier_summary.tex}

Two coupled budget mechanics explain why cold restarts simultaneously lower solve rates and increase mean token consumption under a shared cap $B$. First, terminating a run at landmark $b = 20\text{k}$ restricts the replacement attempt to a residual budget of $B - b$ tokens ($R_j = Y_j\mathbf{1}(L_j \le B-b)$), whereas continuation permits the initial run to utilize the full cap $B$. At tighter caps ($B = 30\text{k}$ and $50\text{k}$, leaving $10\text{k}$ and $30\text{k}$ residual tokens), Fig.~\ref{fig:landmarks} (right) shows that $91\%\text{--}100\%$ of coding trajectories require more than $B - b$ tokens to finish, rendering cold replacements nearly guaranteed to truncate prior to patch submission. Even at $B = 75\text{k}$ and $100\text{k}$, replacement trajectories that would have succeeded between $B - b$ and $B$ tokens are forfeited. Second, aborting an active run at $b$ forfeits the initial $20\text{k}$ tokens invested in repository exploration and forces the replacement attempt to repeat setup and localization from scratch, driving combined token expenditure $K_j = b + \min(L_j, B-b)$ above the cost of allowing moderate-length original attempts to finish naturally.

\rqanswer{RQ3}{Under a shared $75\text{k}$-token cap, learned stopping outperforms matched random restart by up to $1.98$ percentage points through between-task stopping selection, while attempt-level selection contributes minimally. Against uninterrupted continuation, however, cold restarts achieve lower solve rates at higher mean token costs across the $75\text{k}$ and $100\text{k}$ frontiers ($111$ of $112$ configurations, and $216$ of $224$ configurations across all tested caps), showing that improvements over random restart do not establish compute savings.}

\section{Discussion}

Evaluating runtime interventions in autonomous software engineering requires connecting task-level difficulty with attempt-level discrimination and budget-constrained policy mechanics. Our findings explain why strong cross-task ranking does not translate into effective cold restarts under shared token budgets.

\paragraph{The timing bottleneck of trajectory-level signals}
Comparing early prefixes with later checkpoints clarifies why attempt-level discrimination is difficult in repository-level coding. On the primary corpus, $87.51\%$ of active runs have already executed a detected test command and $32.80\%$ have a detected edit action by $20\text{k}$ transcript tokens. Thus, the near-chance within-task discrimination observed at $20\text{k}$ tokens ($\mathrm{AUC}_{\mathrm{within}} \approx 0.50$) cannot be explained simply by an absence of early test execution. Rather, detecting an early test invocation or edit action does not establish whether the interaction was diagnostic or whether its outcome distinguished eventual task resolution from failure. Measurable within-task separation emerges only at later landmarks, rising from $0.52$ at $30\text{k}$ tokens and $0.54$ at $40\text{k}$ tokens to $0.56$ at $60\text{k}$ tokens, and at run completion, where reference-hunk recall reaches a within-task AUC of $0.62$. By those later stages, however, either a substantial fraction of the token budget has been spent or most runs have already terminated. Early failure prediction therefore faces a fundamental timing tension: at landmarks early enough to leave ample residual budget for a replacement attempt, observed prefix signals do not separate competing runs on the same issue.

\paragraph{Why task-level selection confounds stopping baselines}
When a monitor primarily captures static issue difficulty, score thresholding acts as an implicit task filter. Under a shared cap, learned stopping concentrates aborts on hard tasks where continuation rarely succeeds, whereas matched random stopping aborts easy and hard tasks alike, forfeiting impending successes on tractable issues ($\Delta_{\mathrm{between}} > 0$) despite near-chance attempt ranking ($\mathrm{AUC}_{\mathrm{within}} \approx 0.50$). For the same reason, mid-run \emph{abort-only} termination wastes $20\text{k}$ tokens executing a task filter that could be applied pre-execution at zero token cost. Against uninterrupted continuation, forfeiting initial exploration while compressing the residual budget causes cold restarts to underperform across the budget frontier.

\paragraph{Designing compute-efficient agent interventions}
Separating task difficulty from attempt progress points to two compute-management strategies:

\noindent\textbf{1) Pre-execution routing and budgeting:} Because static signals rank difficulty before execution, they naturally support up-front token provisioning~\cite{disentangling} and solver routing~\cite{scrouting}. On shared Qwen3.5 tasks, scaffold choice alone shifts solve rates by $14.06$ percentage points (SWE-agent $47.30\%$ vs.\ OpenHands $33.24\%$), dwarfing the at most $1.98$ percentage-point gain of mid-run restarts and motivating pre-execution routing across frameworks.

\noindent\textbf{2) State-preserving recovery:} Because cold prefixes lack attempt-level signal, retaining intermediate edits and test diagnostics is essential to avoid repeating exploration, explaining why state-preserving recovery substantially outperforms cold restarts~\cite{failfast} and motivating cross-framework context transfer when runs stall.

\paragraph{Evaluation standards for mining agent traces}
As public corpora of agent trajectories grow~\cite{nebius,nvidia}, runtime monitoring studies should evaluate both attempt-level discrimination and net compute efficiency. Specifically, predictive evaluations should report within-task AUC ($\mathrm{AUC}_{\mathrm{within}}$) across repeated attempts alongside pre-execution issue baselines and permutation controls. Policy evaluations under shared token budgets should compare proposed stopping rules against both matched random stopping and uninterrupted continuation, decomposing outcomes into between-task and within-task selection to verify genuine attempt-level value.

\section{Threats to Validity}

\paragraph{Internal validity}
Because our study mines public execution corpora~\cite{nebius,nvidia}, outcome labels and trajectory contents reflect the original benchmark harnesses and execution environments. To guard against data-cleaning artifacts or estimation bias, we audited source task identifiers for text consistency and evaluated within-task label-permutation controls across all conditional modeling pipelines; permutation controls centered at chance ($0.50$) in the primary and OpenHands settings and identified the calibration deviation in SWE-agent/Qwen3.5, where we refrained from formal equivalence claims.

\paragraph{Construct validity}
We measure computational progress in transcript tokens to standardize comparisons across corpora; however, transcript length abstracts away provider-level caching discounts and wall-clock tool execution latency. Our neural monitor adapts the single-landmark terminal-outcome classifier from FailFast--RestartSmart~\cite{failfast}, omitting intermediate test supervision and state-preserving edit transfer to isolate the predictive signal of raw execution prefixes and the mechanics of cold restarts; additionally, the editor-tool heuristic used to include an earlier prefix message can match file-view operations alongside code modifications. Recorded-attempt replay evaluates binary task resolution under hard token caps, assigning zero credit to partial edits that exceed the residual budget, and released task-category labels in Open-SWE-Traces may reflect post-hoc curation rather than metadata available at issue creation.

\paragraph{External validity}
Our findings are scoped to four open-weight model-framework settings on Python-centric repository benchmarks evaluated with linear count and embedding classifiers alongside a $0.6\text{B}$-parameter neural monitor. Within-task discrimination requires mixed-outcome tasks by definition. Monitors that probe internal model representations~\cite{doomed,horizons} or incorporate sequential verifier feedback and state-preserving recovery~\cite{evaluator,failfast} may exhibit different discrimination profiles.

\paragraph{Conclusion validity}
Cluster-bootstrap confidence intervals resample out-of-fold predictions at the cluster level rather than refitting models per bootstrap draw. Subgroup and multi-landmark sensitivity checks apply Holm--Bonferroni multiplicity adjustments~\cite{holm1979}, whereas confidence intervals at the primary $20\text{k}$ landmark are reported unadjusted. Finally, Pareto dominance across the $224$ budget-threshold configurations is evaluated on empirical replay point estimates.

\section{Conclusion}
Early failure prediction benefits autonomous coding agents only when it helps decide whether to abandon an active attempt, rather than merely identifying which software issues are difficult. Mining $147,002$ execution traces across four model-framework settings, we find that early execution prefixes add minimal predictive signal beyond static task context. Across early decision points, initial-turn frozen embeddings ($5\text{k}$ tokens) as well as $20\text{k}$-token action-count models, task-conditional rankers, and a three-seed LoRA-adapted Qwen3-0.6B monitor ($32,406$ attempts) all yield within-task discrimination near chance ($\mathrm{AUC}_{\mathrm{within}} \approx 0.50$), including across high-solvability tasks.

Evaluating stopping rules through budget-constrained replay shows why predictive ranking and policy efficiency must be assessed together. Under a shared $75\text{k}$-token cap, learned stopping outperforms matched random restart by up to $1.98$ percentage points in two settings, but exact decomposition reveals that this gain is driven by between-task stopping allocation rather than within-task failure detection. More importantly, uninterrupted continuation achieves higher solve rates at lower mean token cost than cold restarts across the $75\text{k}$ and $100\text{k}$ frontiers ($111$ of $112$ configurations, and $216$ of $224$ configurations across all tested caps). Future early-stopping and recovery methods should therefore evaluate discrimination within repeated attempts on the same task and demonstrate net compute savings relative to uninterrupted continuation.


\bibliographystyle{IEEEtran}
\bibliography{references}
\end{document}

%% file: tables/data.tex
\begin{table}[t]
\centering\small
\setlength{\tabcolsep}{3pt}
\caption{Summary of released execution-trace settings. OH: OpenHands; SA: SWE-agent. ICC measures within-task outcome clustering on the binary scale.}\label{tab:data}
\begin{tabular}{lrrrr}\toprule
Setting & Runs & Tasks & Repositories & ICC \\
\midrule
OH / Qwen3-Coder & 67,074 & 6,306 & 1,823 & 0.797 \\
SA / Qwen3.5 & 15,098 & 7,885 & 1,815 & 0.768 \\
OH / Qwen3.5 & 30,957 & 14,142 & 2,234 & 0.713 \\
OH / MiniMax-M2.5 & 33,873 & 14,716 & 2,275 & 0.815 \\
\bottomrule\end{tabular}
\end{table}

%% file: figures/landmarks.tex
\begin{tikzpicture}[baseline=(current axis.north)]
\begin{axis}[font=\footnotesize,grid=major,grid style={gray!15},tick align=outside,scaled ticks=false,width=0.46\textwidth,height=4.9cm,xmin=3,xmax=62,xtick={5,20,40,60},xlabel={Transcript tokens (thousands)},ymin=.40,ymax=.64,ylabel={Within-task AUC},title={(a) Ranking within tasks},legend style={font=\scriptsize,at={(.5,-.35)},anchor=north,legend columns=2,draw=none}]
\addplot[black,dashed,forget plot] coordinates {(3,.5)(62,.5)};
\addplot[color=blue!75!black,mark=*,mark size=1.5pt,thick,error bars/.cd,y dir=both,y explicit] coordinates {(5.0,0.4971096018249689) += (0,0.010930790127188117) -= (0,0.011598199695016354) (10.0,0.49386924512650354) += (0,0.010563609135375018) -= (0,0.010050082864980991) (15.0,0.4918196432275462) += (0,0.01055710228740997) -= (0,0.010480688247588632) (20.0,0.49384867479907407) += (0,0.011770988487031453) -= (0,0.011647393107510728) (30.0,0.5168900362517851) += (0,0.012508283019366573) -= (0,0.012973648585570596) (40.0,0.5366580640471085) += (0,0.014078319636336656) -= (0,0.014904907861971961) (60.0,0.5571851642927654) += (0,0.027243088961946826) -= (0,0.02654321621024125)};
\addlegendentry{OH / Qwen3-Coder}
\addplot[color=orange!85!black,mark=square*,mark size=1.5pt,thick,error bars/.cd,y dir=both,y explicit] coordinates {(5.0,0.513051305130513) += (0,0.03341617002584851) -= (0,0.03298198473189501) (10.0,0.49324932493249324) += (0,0.03163447026065164) -= (0,0.03297225767318429) (15.0,0.4900990099009901) += (0,0.035386021335629736) -= (0,0.029441848665998638) (20.0,0.49054905490549056) += (0,0.033680386474701784) -= (0,0.030388489165150356) (30.0,0.48558558558558557) += (0,0.029335609898961945) -= (0,0.03206221142653776) (40.0,0.4945652173913043) += (0,0.0317573354429439) -= (0,0.03657877504806728) (60.0,0.5055679287305123) += (0,0.03652765950478176) -= (0,0.03700022406656972)};
\addlegendentry{SA / Qwen3.5}
\addplot[color=teal!80!black,mark=triangle*,mark size=1.5pt,thick,error bars/.cd,y dir=both,y explicit] coordinates {(5.0,0.4886164623467601) += (0,0.01997688649705004) -= (0,0.019885479526800642) (10.0,0.4903677758318739) += (0,0.020268121703743536) -= (0,0.019487898766613887) (15.0,0.49106830122591943) += (0,0.019511070590718937) -= (0,0.021882045698354335) (20.0,0.489492119089317) += (0,0.020260836255254655) -= (0,0.020323314059922215) (30.0,0.47702560505085934) += (0,0.02035512055468358) -= (0,0.02043803184667836) (40.0,0.4998144712430427) += (0,0.022121708107406435) -= (0,0.01999485049532268) (60.0,0.5304740406320542) += (0,0.031210415003857794) -= (0,0.03278114204910321)};
\addlegendentry{OH / Qwen3.5}
\addplot[color=purple,mark=diamond*,mark size=1.5pt,thick,error bars/.cd,y dir=both,y explicit] coordinates {(5.0,0.48469176665287544) += (0,0.018284857360889528) -= (0,0.017447245921923593) (10.0,0.48407116259826233) += (0,0.02103754319201545) -= (0,0.02265250926874346) (15.0,0.4985519238725693) += (0,0.02310833794888023) -= (0,0.020844685042432087) (20.0,0.4868859283930058) += (0,0.022565094660080054) -= (0,0.022710059448372788) (30.0,0.4735159817351598) += (0,0.022743038698989193) -= (0,0.022687392234843506) (40.0,0.47493887530562345) += (0,0.02552566419772745) -= (0,0.023863325117754464) (60.0,0.4895397489539749) += (0,0.04901042083816487) -= (0,0.051117688009149786)};
\addlegendentry{OH / MiniMax}
\end{axis}
\end{tikzpicture}
\hfill
\begin{tikzpicture}[baseline=(current axis.north)]
\begin{axis}[font=\footnotesize,grid=major,grid style={gray!15},tick align=outside,scaled ticks=false,width=0.46\textwidth,height=4.9cm,xmin=3,xmax=62,xtick={5,20,40,60},xlabel={Transcript tokens (thousands)},ymin=0,ymax=1.05,ylabel={Fraction still active},title={(b) Remaining population},legend style={font=\scriptsize,at={(.5,-.35)},anchor=north,legend columns=2,draw=none}]
\addplot[color=blue!75!black,mark=*,mark size=1.5pt,thick,] coordinates {(5.0,1.0) (10.0,1.0) (15.0,0.9998509109341921) (20.0,0.9974953036944271) (30.0,0.9270656290067687) (40.0,0.6926976175567284) (60.0,0.22794227271371917)};
\addplot[color=orange!85!black,mark=square*,mark size=1.5pt,thick,] coordinates {(5.0,1.0) (10.0,1.0) (15.0,0.9998012981851901) (20.0,0.9998012981851901) (30.0,0.9987415551728706) (40.0,0.9866207444694661) (60.0,0.7995098688568022)};
\addplot[color=teal!80!black,mark=triangle*,mark size=1.5pt,thick,] coordinates {(5.0,1.0) (10.0,1.0) (15.0,1.0) (20.0,0.9999676971282747) (30.0,0.9977387989792292) (40.0,0.9634331492069645) (60.0,0.5840036179216332)};
\addplot[color=purple,mark=diamond*,mark size=1.5pt,thick,] coordinates {(5.0,1.0) (10.0,0.9999114338853955) (15.0,0.9979629793640953) (20.0,0.9904643816609099) (30.0,0.9129395093437251) (40.0,0.7179169249845009) (60.0,0.28612759424910694)};
\end{axis}
\end{tikzpicture}

%% file: figures/pooled_within.tex
\begin{tikzpicture}
\begin{axis}[font=\footnotesize,grid=major,grid style={gray!15},tick align=outside,scaled ticks=false,width=.86\columnwidth,height=4.6cm,xmin=.38,xmax=1.02,ymin=-.6,ymax=3.6,xtick={.4,.5,.6,.7,.8,.9,1},xticklabel style={/pgf/number format/fixed},ytick={0,1,2,3},yticklabels={Counts,Embeddings,Task rate,Task rate + counts},xlabel={AUC at 5k tokens (primary corpus)},legend style={at={(.5,-.38)},anchor=north,draw=none,font=\scriptsize},legend columns=2]
\addplot[black,dashed,forget plot] coordinates {(.5,-.6)(.5,3.6)};
\addplot[gray!60,thick,forget plot] coordinates {(0.4971,0)(0.5772,0)};
\addplot[gray!60,thick,forget plot] coordinates {(0.4922,1)(0.6226,1)};
\addplot[gray!60,thick,forget plot] coordinates {(0.5,2)(0.9657,2)};
\addplot[gray!60,thick,forget plot] coordinates {(0.5008,3)(0.9662,3)};
\addplot[only marks,color=blue!75!black,mark=*,mark size=2pt] coordinates {(0.5772,0)(0.6226,1)(0.9657,2)(0.9662,3)};
\addlegendentry{Pooled AUC}
\addplot[only marks,color=orange!85!black,mark=o,mark size=2pt,thick] coordinates {(0.4971,0)(0.4922,1)(0.5,2)(0.5008,3)};
\addlegendentry{Within-task AUC}
\node[font=\scriptsize,anchor=south] at (axis cs:0.5772,0.12) {0.577};
\node[font=\scriptsize,anchor=south] at (axis cs:0.6226,1.12) {0.623};
\node[font=\scriptsize,anchor=south] at (axis cs:0.9657,2.12) {0.966};
\node[font=\scriptsize,anchor=south] at (axis cs:0.9662,3.12) {0.966};
\node[font=\scriptsize,anchor=east] at (axis cs:0.4871,0) {0.497};
\node[font=\scriptsize,anchor=east] at (axis cs:0.4822,1) {0.492};
\node[font=\scriptsize,anchor=east] at (axis cs:0.4900,2) {0.500};
\node[font=\scriptsize,anchor=east] at (axis cs:0.4908,3) {0.501};
\end{axis}
\end{tikzpicture}

%% file: tables/neural_large.tex
\begin{table}[t]
\centering\small
\setlength{\tabcolsep}{2.5pt}
\caption{Three-seed LoRA-adapted Qwen3-0.6B evaluation at 20k tokens across three repository-disjoint folds (311, 285, and 263 mixed-outcome tasks). Metrics are averaged across seeds rather than pooled across seed scores; brackets show 95\% repository-and-seed bootstrap intervals for $\mathrm{AUC}_{\mathrm{within}}$. Lower log loss is better.}\label{tab:neural-large}
\begin{tabular}{llccc}\toprule
Fold & Input & Pooled AUC & Within-task AUC & Log loss \\
\midrule
1 & Issue & 0.526 & 0.500 [0.499, 0.501] & 0.693 \\
1 & Issue + prefix & 0.523 & 0.500 [0.476, 0.526] & 1.788 \\
2 & Issue & 0.513 & 0.500 [0.499, 0.500] & 0.693 \\
2 & Issue + prefix & 0.504 & 0.496 [0.475, 0.517] & 1.012 \\
3 & Issue & 0.527 & 0.500 [0.497, 0.501] & 0.693 \\
3 & Issue + prefix & 0.469 & 0.490 [0.467, 0.513] & 0.879 \\
\bottomrule\end{tabular}
\end{table}

%% file: tables/pass_strata.tex
\begin{table}[t]
\centering
\caption{Within-task AUC at 20k tokens stratified by empirical task solve rate $p$: low ($p<0.25$), medium ($0.25\leq p\leq0.75$), and high ($p>0.75$). Count denotes the primary-corpus issue-plus-prefix count model (95\% task-bootstrap CIs); Neural denotes the three-seed Qwen3-0.6B mean (95\% repository-and-seed CIs). All intervals contain $0.50$. Strata are descriptive post-hoc tiers.}\label{tab:pass-strata}
\setlength{\tabcolsep}{3pt}
\begin{tabular}{llrrl}
\toprule
Predictor & Stratum & Tasks & Pairs & AUC [95\% interval] \\
\midrule
Count & Low-pass & 373 & 5,545 & 0.513 [0.484, 0.541] \\
Count & Medium-pass & 800 & 23,378 & 0.488 [0.476, 0.502] \\
Count & High-pass & 564 & 9,524 & 0.496 [0.475, 0.519] \\
\midrule
Neural & Low-pass & 170 & 2,188 & 0.504 [0.464, 0.544] \\
Neural & Medium-pass & 405 & 10,966 & 0.491 [0.473, 0.509] \\
Neural & High-pass & 284 & 3,664 & 0.503 [0.475, 0.532] \\
\bottomrule
\end{tabular}
\end{table}

%% file: tables/restart.tex
\begin{table}[t]
\centering
\caption{Restart replay at the 20k-token landmark using the median eligible-score threshold. Solve rates are percentages; $\Delta$ is learned minus matched random restart in percentage points, with 95\% repository-bootstrap confidence intervals.}\label{tab:restart}
\scriptsize
\setlength{\tabcolsep}{3pt}
\resizebox{\columnwidth}{!}{%
\begin{tabular}{lrrl}
\toprule
Setting & Continue & Learned & $\Delta$ [95\% interval] \\
\midrule
OH / Qwen3-Coder & 46.15 & 41.53 & $-0.458$ [$-0.677$, $-0.219$] \\
SA / Qwen3.5 & 28.71 & 20.66 & $+1.984$ [$+1.313$, $+2.600$] \\
OH / Qwen3.5 & 27.19 & 20.16 & $+0.768$ [$+0.302$, $+1.221$] \\
OH / MiniMax-M2.5 & 41.05 & 35.76 & $-0.663$ [$-1.014$, $-0.333$] \\
\bottomrule
\end{tabular}
}
\end{table}

%% file: tables/replay_decomposition.tex
\begin{table}[t]
\centering
\caption{Exact decomposition of learned minus matched random solve rate (in percentage points) at the 20k landmark under a 75k-token cap. Task means center over attempts active at 20k and normalize by all starting attempts. Between-task selection ($\Delta_{\mathrm{between}}$) governs the sign in every setting; components are rounded independently.}\label{tab:replay-decomp}
\setlength{\tabcolsep}{3pt}
\begin{tabular}{lrrr}
\toprule
Setting & Total & Between & Within \\
\midrule
OH / Qwen3-Coder & $-0.458$ & $-0.494$ & $+0.036$ \\
OH / MiniMax-M2.5 & $-0.663$ & $-0.704$ & $+0.042$ \\
OH / Qwen3.5 & $+0.768$ & $+0.881$ & $-0.113$ \\
SA / Qwen3.5 & $+1.984$ & $+1.749$ & $+0.235$ \\
\bottomrule
\end{tabular}
\end{table}

%% file: tables/frontier_summary.tex
\begin{table}[t]
\centering
\caption{Cost-solve comparisons against uninterrupted continuation at the 20k-token landmark (56 restart configurations per cap across seven thresholds, four settings, and learned vs.\ matched random policies). Ties match continuation in both cost and solve rate. The sole non-dominated point is learned restart for OH/MiniMax at 100k (2nd-percentile threshold: $+0.0033$ percentage points at $+376$ mean tokens).}\label{tab:frontier-summary}
\setlength{\tabcolsep}{4pt}
\begin{tabular}{rrrr}
\toprule
Cap & Dominated & Tied & Distinct trade-offs \\
\midrule
30k & 49 & 7 & 0 \\
50k & 56 & 0 & 0 \\
75k & 56 & 0 & 0 \\
100k & 55 & 0 & 1 \\
\bottomrule
\end{tabular}
\end{table}

%% file: references.bib
@misc{beyond,
 author={Tural Mehtiyev and Wesley Assun{\c{c}}{\~a}o},
 title={Beyond Resolution Rates: Behavioral Drivers of Coding Agent Success and Failure},
 year={2026}, howpublished={arXiv:2604.02547v1}, url={https://arxiv.org/abs/2604.02547v1}}

@misc{agentstop,
 author={Dzung Pham and Kleomenis Katevas and Ali Shahin Shamsabadi and Hamed Haddadi},
 title={{AgentStop}: Terminating Local {AI} Agents Early to Save Energy in Consumer Devices},
 year={2026}, howpublished={arXiv:2605.15206v1}, url={https://arxiv.org/abs/2605.15206v1}}

@misc{automata,
 author={Seonglae Cho and Franklin Cardenoso Fernandez and Umar Mohammed and Zekun Wu and Kleyton Da Costa and Ilham Wicaksono and Adriano Koshiyama},
 title={Automata from Agent Traces: Failure and Next-Step Prediction},
 year={2026}, howpublished={arXiv:2608.23670v1}, url={https://arxiv.org/abs/2608.23670v1}}

@misc{consistency,
 author={Amritesh Banerjee and Pranil Raichura},
 title={Measuring Cross-Task Behavioral Consistency in Language Model Agents},
 year={2026}, howpublished={arXiv:2608.13598v1}, url={https://arxiv.org/abs/2608.13598v1}}

@misc{sparse,
 author={Avinash Baidya and Xinran Liang and Ruocheng Guo and Xiang Gao and Kamalika Das},
 title={When Evidence is Sparse: Weakly Supervised Early Failure Alerting in Dialogs and {LLM}-Agent Trajectories},
 year={2026}, howpublished={arXiv:2606.05414v1}, url={https://arxiv.org/abs/2606.05414v1}}

@misc{horizons,
 author={Andr{\'e} Silva and Han Tu and Martin Monperrus},
 title={Latent Programming Horizons in Coding Agents},
 year={2026}, howpublished={arXiv:2607.05188v1}, url={https://arxiv.org/abs/2607.05188v1}}

@misc{scrouting,
 author={Ishaan Bhola and Adithyan Krishnan and {Mukunda NS}},
 title={Scrouting: Cost-Aware Routing of Coding Agents by Scouting the Repository First},
 year={2026}, howpublished={arXiv:2608.04804v1}, url={https://arxiv.org/abs/2608.04804v1}}

@misc{nebius,
 author={{Nebius}}, title={{SWE-rebench OpenHands Trajectories}},
 howpublished={Hugging Face dataset}, year={2026},
 url={https://huggingface.co/datasets/nebius/SWE-rebench-openhands-trajectories}}

@misc{nvidia,
 author={{NVIDIA}}, title={{Open-SWE-Traces}},
 howpublished={Hugging Face dataset}, year={2026},
 url={https://huggingface.co/datasets/nvidia/Open-SWE-Traces}}

@misc{failfast,
 author={Chenyu Wang and Yunbo Lyu and Junda He and Zhou Yang and Chenxing Zhong and Yaniv Harel and David Lo},
 title={Fail-Fast, Restart-Smart: Early Failure Prediction and Restart for {SWE} Agentic Tasks},
 year={2026}, howpublished={arXiv:2608.03222v1}, url={https://arxiv.org/abs/2608.03222v1}}

@misc{eet,
 author={Yaoqi Guo and Ying Xiao and Jie M. Zhang and Mark Harman and Yiling Lou and Yang Liu and Zhenpeng Chen},
 title={{EET}: Experience-Driven Early Termination for Cost-Efficient Software Engineering Agents},
 year={2026}, howpublished={arXiv:2601.05777v2}, url={https://arxiv.org/abs/2601.05777v2}}

@misc{bagen,
 author={Yuxiang Lin and Zihan Wang and Mengyang Liu and Yuxuan Shan and Longju Bai and Junyao Zhang and Xing Jin and Boshan Chen and Jinyan Su and Xingyao Wang and Jiaxin Pei and Manling Li},
 title={{BAGEN}: Are {LLM} Agents Budget-Aware?},
 year={2026}, howpublished={arXiv:2606.00198v1}, url={https://arxiv.org/abs/2606.00198v1}}

@misc{evaluator,
 author={Shuvom Sadhuka and Drew Prinster and Clara Fannjiang and Gabriele Scalia and Bonnie Berger and Aviv Regev and Hanchen Wang},
 title={E-valuator: Reliable Agent Verifiers with Sequential Hypothesis Testing},
 year={2026}, howpublished={arXiv:2512.03109v2}, url={https://arxiv.org/abs/2512.03109v2}}

@misc{process,
 author={Xiangxin Zhao and Han Li and Shuaiting Li and Tianyi Zhao and Earl T. Barr and Federica Sarro and He Ye},
 title={Failure as a Process: An Anatomy of {CLI} Coding Agent Trajectories},
 year={2026}, howpublished={arXiv:2607.09510v1}, url={https://arxiv.org/abs/2607.09510v1}}

@misc{doomed,
 author={Kai Ruan and Zihe Huang and Ziqi Zhou and Qianshan Wei and Jinghao Lin and Xuan Wang and Hao Sun},
 title={Doomed from the Start: Early Abort of {LLM} Agent Episodes via a Recall-Controlled Probe Cascade},
 year={2026}, howpublished={arXiv:2607.06503v2}, url={https://arxiv.org/abs/2607.06503v2}}

@inproceedings{swebench,
  title={Swe-bench: Can language models resolve real-world github issues?},
  author={Jimenez, Carlos E and Yang, John and Wettig, Alexander and Yao, Shunyu and Pei, Kexin and Press, Ofir and Narasimhan, Karthik},
  booktitle={International Conference on Learning Representations},
  volume={2024},
  pages={54107--54157},
  year={2024}
}

@article{sweagent,
  title={Swe-agent: Agent-computer interfaces enable automated software engineering},
  author={Yang, John and Jimenez, Carlos and Wettig, Alexander and Lieret, Kilian and Yao, Shunyu and Narasimhan, Karthik and Press, Ofir},
  journal={Advances in Neural Information Processing Systems},
  volume={37},
  pages={50528--50652},
  year={2024}
}

@article{openhands,
  title={Openhands: An open platform for ai software developers as generalist agents, 2024},
  author={Wang, Xingyao and Li, Boxuan and Song, Yufan and Xu, Frank F and Tang, Xiangru and Zhuge, Mingchen and Pan, Jiayi and Song, Yueqi and Li, Bowen and Singh, Jaskirat and others},
  journal={URL https://arxiv. org/abs/2407.16741},
  volume={2},
  number={5},
  year={2025}
}

@article{luby1993,
  title={Optimal speedup of Las Vegas algorithms},
  author={Luby, Michael and Sinclair, Alistair and Zuckerman, David},
  journal={Information Processing Letters},
  volume={47},
  number={4},
  pages={173--180},
  year={1993},
  publisher={Elsevier}
}

@article{swerebench,
  title={Swe-rebench: An automated pipeline for task collection and decontaminated evaluation of software engineering agents},
  author={Badertdinov, Ibragim and Golubev, Alexander and Nekrashevich, Maksim and Shevtsov, Anton and Karasik, Simon and Andriushchenko, Andrei and Trofimova, Maria and Litvintseva, Daria and Yangel, Boris},
  journal={Advances in Neural Information Processing Systems},
  volume={38},
  year={2026}
}

@misc{qwen3coder,
 author={{Qwen Team}}, title={{Qwen3-Coder-480B-A35B-Instruct}},
 howpublished={Hugging Face model card}, year={2025},
 url={https://huggingface.co/Qwen/Qwen3-Coder-480B-A35B-Instruct}}

@misc{qwen35,
 author={{Qwen  Team}}, title={{Qwen3.5-122B-A10B}},
 howpublished={Hugging Face model card}, year={2026},
 url={https://huggingface.co/Qwen/Qwen3.5-122B-A10B}}

@misc{minimaxm25,
 author={{MiniMax}}, title={{MiniMax-M2.5}},
 howpublished={Hugging Face model card}, year={2026},
 url={https://huggingface.co/MiniMaxAI/MiniMax-M2.5}}

@article{schuirmann1987,
  title={A comparison of the two one-sided tests procedure and the power approach for assessing the equivalence of average bioavailability},
  author={Schuirmann, Donald J},
  journal={Journal of pharmacokinetics and biopharmaceutics},
  volume={15},
  number={6},
  pages={657--680},
  year={1987},
  publisher={Springer}
}

@misc{lakens2017,
  title={Equivalence tests: A practical primer for t tests, correlations, and meta-analyses. Social Psychological and Personality Science, 8 (4), 355--362},
  author={Lakens, Dani{\"e}l},
  year={2017}
}

@inproceedings{sbert,
  title={Sentence-bert: Sentence embeddings using siamese bert-networks},
  author={Reimers, Nils and Gurevych, Iryna},
  booktitle={Proceedings of the 2019 conference on empirical methods in natural language processing and the 9th international joint conference on natural language processing (EMNLP-IJCNLP)},
  pages={3982--3992},
  year={2019}
}

@article{minilm,
  title={Minilm: Deep self-attention distillation for task-agnostic compression of pre-trained transformers},
  author={Wang, Wenhui and Wei, Furu and Dong, Li and Bao, Hangbo and Yang, Nan and Zhou, Ming},
  journal={Advances in neural information processing systems},
  volume={33},
  pages={5776--5788},
  year={2020}
}

@article{qwen3,
  title={Qwen3 technical report},
  author={Yang, An and Li, Anfeng and Yang, Baosong and Zhang, Beichen and Hui, Binyuan and Zheng, Bo and Yu, Bowen and Gao, Chang and Huang, Chengen and Lv, Chenxu and others},
  journal={arXiv preprint arXiv:2505.09388},
  year={2025}
}

@article{lora,
  title={Lora: Low-rank adaptation of large language models},
  author={Hu, Edward J and Shen, Yelong and Wallis, Phillip and Allen-Zhu, Zeyuan and Li, Yuanzhi and Wang, Shean and Wang, Lu and Chen, Weizhu},
  journal={arXiv preprint arXiv:2106.09685},
  year={2021}
}

@article{hanley1982,
  title={The meaning and use of the area under a receiver operating characteristic (ROC) curve.},
  author={Hanley, James A and McNeil, Barbara J},
  journal={Radiology},
  volume={143},
  number={1},
  pages={29--36},
  year={1982}
}

@article{field2007,
  title={Bootstrapping clustered data},
  author={Field, Christopher A and Welsh, Alan H},
  journal={Journal of the Royal Statistical Society Series B: Statistical Methodology},
  volume={69},
  number={3},
  pages={369--390},
  year={2007},
  publisher={Oxford University Press}
}

@article{ridout1999,
  title={Estimating intraclass correlation for binary data},
  author={Ridout, Martin S and Dem{\'e}trio, Clarice GB and Firth, David},
  journal={Biometrics},
  volume={55},
  number={1},
  pages={137--148},
  year={1999},
  publisher={Oxford University Press}
}

@article{holm1979,
  title={A simple sequentially rejective multiple test procedure},
  author={Holm, Sture},
  journal={Scandinavian journal of statistics},
  pages={65--70},
  year={1979},
  publisher={JSTOR}
}

@inproceedings{robbes2026,
  title={Promises, perils, and (timely) heuristics for mining coding agent activity},
  author={Robbes, Romain and Matricon, Th{\'e}o and Degueule, Thomas and Hora, Andre and Zacchiroli, Stefano},
  booktitle={Proceedings of the 23rd International Conference on Mining Software Repositories},
  pages={496--507},
  year={2026}
}

@misc{disentangling,
  author       = {EsfandyariDoulabi, Mohsen and Arkoh, Lawrence and Tadesse, Biruk and Patel, Vaishvi and Sharma, Mehul and d'Amorim, Marcelo and Assun{\c{c}}{\~a}o, Wesley},
  title        = {Disentangling Task Difficulty from Run-Level Failure in Agent Failure Prediction},
  howpublished = {arXiv:2610.05572v1},
  year         = {2026},
  url          = {https://arxiv.org/abs/2610.05572v1}
}

@misc{problempath,
  author       = {Bulut, Yigit Utku},
  title        = {It's the Problem, Not the Path: Budget and Difficulty Confounds in {LLM} Reasoning Trajectories},
  howpublished = {arXiv:2609.03436v1},
  year         = {2026},
  url          = {https://arxiv.org/abs/2609.03436v1}
}

@article{breslow1980statistical,
  title={Statistical methods in cancer research},
  author={Breslow, Norman E and Day, Nicholas E and Heseltine, Elisabeth},
  year={1980},
  publisher={International agency for research on cancer, Lyon}
}

@inproceedings{sculley2010combined,
  title={Combined regression and ranking},
  author={Sculley, David},
  booktitle={Proceedings of the 16th ACM SIGKDD international conference on Knowledge discovery and data mining},
  pages={979--988},
  year={2010}
}

@article{kitagawa1955components,
  title={Components of a difference between two rates},
  author={Kitagawa, Evelyn M},
  journal={Journal of the american statistical association},
  volume={50},
  number={272},
  pages={1168--1194},
  year={1955},
  publisher={Taylor \& Francis}
}

@article{gneiting2007strictly,
  title={Strictly proper scoring rules, prediction, and estimation},
  author={Gneiting, Tilmann and Raftery, Adrian E},
  journal={Journal of the American statistical Association},
  volume={102},
  number={477},
  pages={359--378},
  year={2007},
  publisher={Taylor \& Francis}
}
